\documentclass[%
reprint,pra,
superscriptaddress,
amsmath,amssymb,
aps,
]{revtex4}

\usepackage{amssymb}

\usepackage{lineno}
\usepackage{latexsym}
\usepackage{amsmath}
\usepackage{amssymb}
\usepackage{graphicx}
\usepackage{caption}
\usepackage{subfigure}
\usepackage{float}
\usepackage{mathrsfs}
\usepackage{color}
\usepackage[justification=centering,format=plain]{caption}

\begin{document}
	
\preprint{APS/123-QED}

\title{Computational ghost imaging of moving objects with random speed}

\author{Xiaoyu Nie}
\affiliation{%
Texas A\&M University, College Station, Texas 77843, USA}%
\affiliation{%
	Xi'an Jiaotong University, Xi'an, Shaanxi 710049, China}%
\author{Xingchen Zhao}%
\affiliation{%
	Texas A\&M University, College Station, Texas 77843, USA}%
\author{Tao Peng}%
\email{taopeng@tamu.edu}
\affiliation{%
	Texas A\&M University, College Station, Texas 77843, USA}%
\author{Marlan O. Scully}%
\affiliation{%
	Texas A\&M University, College Station, Texas 77843, USA}%
\affiliation{%
	Baylor University, Waco 76706, USA}%
\date{\today}

\begin{abstract}
We show that a recently developed computational ghost imaging method based on the pink noise speckle pattern has a strong ability to image moving objects. To examine our scheme's unique ability and application scope, we experimentally image an object moving with the random speed at variable amplitudes. We show that the pink noise speckle pattern can effectively image moving objects when the traditional methods fail. The results provide promising potential applications for ghost imaging in remote sensing.
\end{abstract}

\maketitle

\section{Introduction}
Ghost imaging (GI)~\cite{Pittman1995Optical,bennink2002two,chen2009lensless}, a practice with single-pixel imaging, possesses advanced features comparing to the conventional photographing. Since it applies the second-order correlation algorithm to photo reconstruction, the GI method can be immune to the media's interference and strong environmental noise. To further ameliorate and simplify the system, computational ghost imaging (CGI)~\cite{bromberg2009ghost,shapiro2008computational} is proposed, where one light path usually recording speckles without passing objects is alternated by loading pre-generated patterns in computer memories onto the digital micro-mirror devices (DMD). Thus, CGI uses a single element detector to reconstruct images by sequentially recording the correlation between a set of intensity and their corresponding patterns, which has non-conventional applications such as wide spectrum imaging~\cite{radwell2014single,Yang2017Ghost}, depth mapping~\cite{sun2016single}, and sub-Rayleigh imaging~\cite{kuplicki2016high,chen2017sub,sprigg2016super}.

However, unlike conventional photographing, which only needs to take one snapshot of the object, CGI requires many samplings to reconstruct a high-quality image. The minimum number of sampling is proportionate to the total number of pixels projecting on objects~\cite{Katz2009Compressive}. It is, therefore, natural to require that the object stays steady when illuminated by a sequence of patterns. The image quality is determined by both the exposure time of each frame and the number of frames, \textit{i.e.}, the total integration time. The image could be blurred or even destroyed if the target is moving. Previously, many schemes have been proposed to improve the speed of CGI~\cite{wang2016fast,gong2016three,xu20181000,jiang2019scan} or reducing the sampling rate to the sub-Nyquist region~\cite{sun2017russian,lyu2017deep, luo2018orthonormalization,He2018Ghost}. Other effects have been made to image objects with motion, using linear correlation algorithm~\cite{li2011lensless}, working in the Fourier domain~\cite{zhang2013improving}, or using entangled photons~\cite{magana2013compressive}. When the object moves with an unknown constant speed in various motion modes, it has been shown that compensation of the motion can be made therefore retrieve the images~\cite{li2014ghost,li2015ghost,jiao2019motion}. Under the condition that the sampling rate of GI is faster than the refreshing motion rate of the image, a clear GI can be gradually obtained~\cite{sun2019gradual}. However, imaging a moving object in general without any prior knowledge of the motion is still quite challenging. We recently developed a method to customize the speckle patterns for CGI in the spatial frequency domain, \textit{i.e.}, the generation of colored noise speckle patterns~\cite{li2021sub,nie2020noise}. Colored noise generally has non-zero cross-correlation between neighborhood pixels as compared with the white noise. In particular, the pink noise speckle pattern has shown ability to image in a variety of noisy environments~\cite{nie2020noise}. 

This work aims to minimize the sampling number and improve the signal-to-noise ratio (SNR) using the synthesized pink noise speckle patterns, when imaging a moving object with an unknown variable speed. The results were compared to the traditional white noise speckle patterns. The unique ability and scope of its applications are demonstrated by several comparisons where the target moves at a random speed with ranges 0${mm}$, 0.4${mm}$, 0.9${mm}$, and 1.8${mm}$. Our results demonstrate that CGI with pink noise is effective in capturing randomly moving targets, while conventional CGI fails..

\section{Computational ghost imaging of moving object}

\begin{figure}[!hbt]
\captionsetup{justification=raggedright,singlelinecheck=false}
\centering\includegraphics[width=0.75\linewidth]{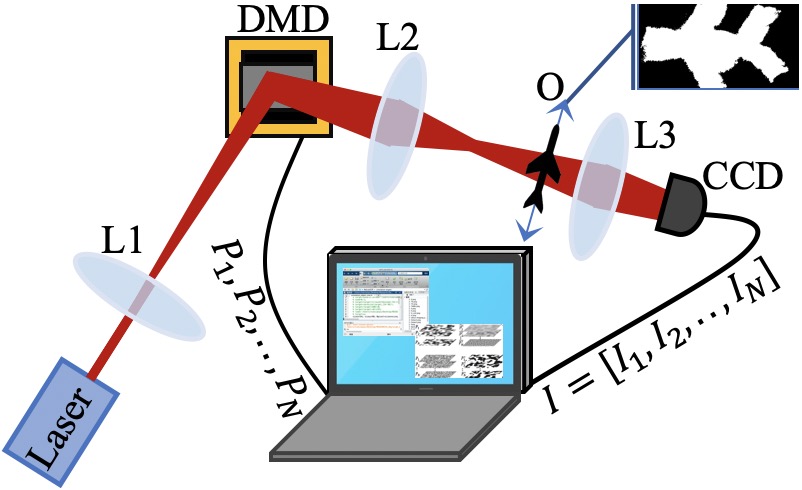}
\caption{Schematic of the setup for a CGI experiment. A CW laser is expended then reflected by a DMD where the noise patterns are loaded. The patterns are imaged by a lens (L2) to the object plane. The \textit{Flight} shaped object is put moves with random speed on a stage. Another lens (L3) is used to image the object plane to the CCD, which can be used as a bucket detector.}
\label{fig:setup}
\end{figure}

The experimental setup is shown in Fig.~\ref{fig:setup}. A CW laser is used to illuminate the DMD. The pattern generated by the DMD is then imaged onto the the target. A lens (L3) is used to image the transmitted light from the target onto the CCD plane. The CCD can be used to monitor the first order image as well as a bucket detector in the CGI system. The DMD contains tiny pixel-mirrors, each measuring 16 $\mu{m}$ by 16 $\mu{m}$. Each 4 by 4 DMD pixels are used as an independently changeable noise speckle pixel.  In our experiment, the size of the noise patterns is 54 by 98 independent pixels. The object is made into a \textit{Flight} shape which contains totally around 2000 independent pixels. In the measurement, The maximum movements of object are made to be  0, 2, 5, and 10 independent pixels on the patterns projected to CCD. The speed and acceleration of object have a random number at every moment. The maximum value of the speed is $500~mm/s$ and the maximum value of the acceleration is $500~mm/s^2 $. The integration time of the CCD is $200~ \mu s$ for each frame. 

\begin{figure}[htbp]
\captionsetup{justification=raggedright,singlelinecheck=false}
\centering\includegraphics[width=0.75\linewidth]{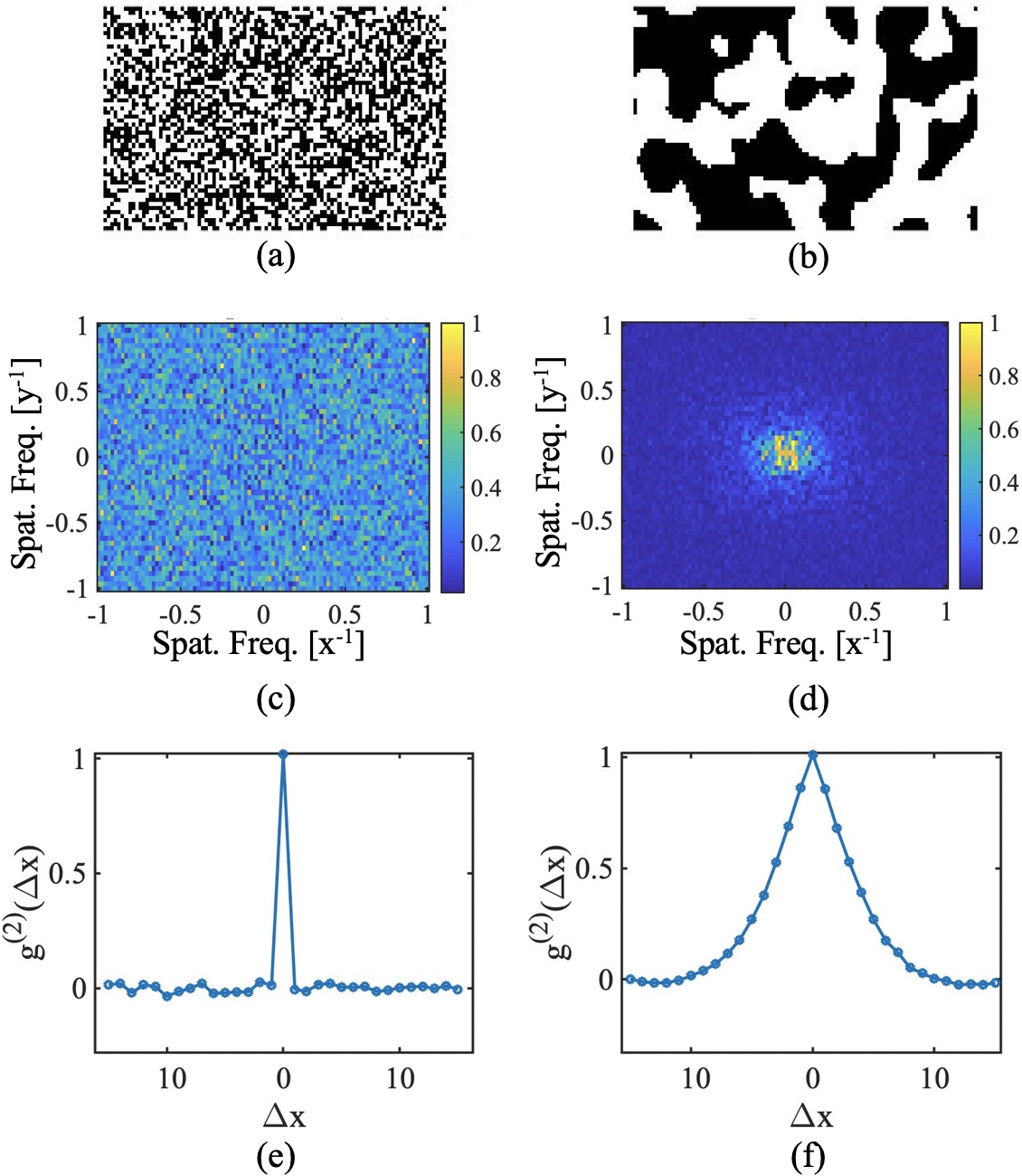}
\caption{A typical two-dimensional noise speckle pattern, its two-dimensional spatial frequency distribution, and one-dimensional cross correlation between pixels for white noise patterns (a), (c), and (e), and pink noise patterns (b), (d), and (f).}
\label{fig:red}
\end{figure}

The noise patterns used in our experiment are binary patterns. For pink noise speckle patterns, the power spectrum density is defined by ${I_{(\omega)}} \sim {\omega ^{-1}}$~\cite{nie2020noise}. Conventional white noise with power spectrum density ${I_{(\omega)}} \sim {\omega ^{0}}$ is also used in our experiments as a comparison. Together with randomly conjugated symmetric phases, an inverse Fourier transformation on the spectrum defined above creates gray-scale ([0, 255]) pink noise patterns. The gray-scale patterns are then converted to binary patterns before applied to the DMD, by defining the pixels over 127.5 to be 1, and pixels below 127.5 to 0. The 2D speckle patterns of white and pink noise are shown in Fig.~\ref{fig:red}(a) and (b). The 2D spatial frequency distributions of white and pink noise are shown in Fig.~\ref{fig:red}(c) and (d). We choose one pixel randomly from the pink noise patterns and calculate its auto-correlation and cross-correlation. Pink noise's spectral intensity distribution leads to a significantly positive correlation between pixels close to each other, as shown in Fig.~\ref{fig:red}(e) and (f). This unique positive cross-correlation is why the pink noise CGI can be employed with reconstructing moving objects, as shown below. 

In CGI algorithms, we should collect a series of light intensity while sampling. The value of each proportionate to the transmitted pattern distribution at the sample position. The light intensity at the bucket detector is
\begin{align}\
I_{\mathrm{i}}\sim \sum\limits_{\widetilde x,\widetilde y}{T(\widetilde x,\widetilde y) P_{\mathrm{i}}(\widetilde x,\widetilde y)},
\end{align}
Where $T$ represents the object transmission function, $P_{\mathrm{i}}$ is the $i$th noise pattern, ${(\widetilde x,\widetilde y)}$ is the pixel notation of patterns on the object plane. For ${(x,y)=(\widetilde x,\widetilde y)}$, the intensity of ${P(\widetilde x,\widetilde y)}$ approximately proportionate to ${P(x,y)}$, weakened due to light propagation. Using the estimated light intensity, we get

\begin{align}
g^{(2)}_{p}(x,y)&= \left\langle P(x,y)I\right\rangle-\left\langle P(x,y)\right\rangle \left\langle I \right\rangle\cr
&=\langle \Delta P(x,y) \Delta I\rangle \cr
&=\frac{1}{N}\langle \Delta P_{\mathrm{i}}(x,y) \sum\limits_{\widetilde x,\widetilde y}T(\widetilde x,\widetilde y) \Delta P_{\mathrm{i}} (\widetilde x,\widetilde y)\rangle \cr
&\propto \sum\limits_{\widetilde x,\widetilde y}T(\widetilde x,\widetilde y)\langle\Delta P(x,y)\Delta P(\widetilde x,\widetilde y)\rangle,
\end{align}
where $N$ is the total number of noise patterns, and ${(x,y)}$ is the pixel notation of patterns in the computer. ${\left\langle \Delta P(x,y) \Delta P(\widetilde x,\widetilde y) \right\rangle }$ is the correlation between the pixels at position $(\widetilde x,\widetilde y)$ and $(x,y)$. For commonly used white noise or speckles in CGI, only when ${(\widetilde x,\widetilde y)=(x,y)}$ we get ${\left\langle \Delta P(x,y) \Delta P(\widetilde x,\widetilde y) \right\rangle=1}$, otherwise ${\left\langle \Delta P(x,y) \Delta P(\widetilde x,\widetilde y) \right\rangle=0}$. Therefor, each point ${(x,y)}$ in the reconstructed image is only contributed by the auto-correlation, thus hard to suppress the environmental noise with a limited number of sampling. On the contrary, for the pink noise CGI, in addition to the contribution from terms where ${(\widetilde x,\widetilde y)=(x,y)}$, ${\left\langle \Delta P(x,y) \Delta P(\widetilde x,\widetilde y) \right\rangle=1}$, the term whose ${(\widetilde x,\widetilde y)}$, ${(x,y)}$ are neighbours also make contributions because their cross-correlation ${\left\langle \Delta P(x,y) \Delta P(\widetilde x,\widetilde y) \right\rangle > 0}$. Thus, the intensity of every point ${(x,y)}$ in the reconstructed image is derived by adding up both auto-correlation and cross-correlation. Here, the cross-correlation was defined by ${\left\langle \Delta P(x,y)\Delta P(\widetilde x,\widetilde y) \right\rangle}$.
Consequently, the pink noise CGI can reconstruct the image when the equivalent static sampling number drops if the target moves. Meanwhile, the pink noise CGI can suppress the external noise and enhance the visibility, which can be reflected in the SNR.

\section{Experiment results}

In the experiment, we set the distance of motion as 0${mm}$, 0.4${mm}$, 0.9${mm}$, and 1.8${mm}$, which correspond to 0, 2, 5, and 10 independent changeable pixels of noise patterns projected on the CCD surface, respectively. For each group, pink noise and white noise are applied, and CGI results are achieved from variable sampling numbers ($N = 1000, 2000, 5000, 10000$). Also, we introduce SNR to justify the quality of results. The definition of SNR is 
\begin{equation}
{SNR} = \frac{\mu_{\mathrm{sig}}}{\sigma_{\mathrm{sig}}}.
\end{equation}
where the $\mu_{\mathrm{sig}}$ is the average signal value and the $\sigma_{\mathrm{sig}}$ is the standard deviation of the signal.

\begin{figure}[!htp]
\captionsetup{justification=raggedright,singlelinecheck=false}
\centering\includegraphics[width=0.75\linewidth]{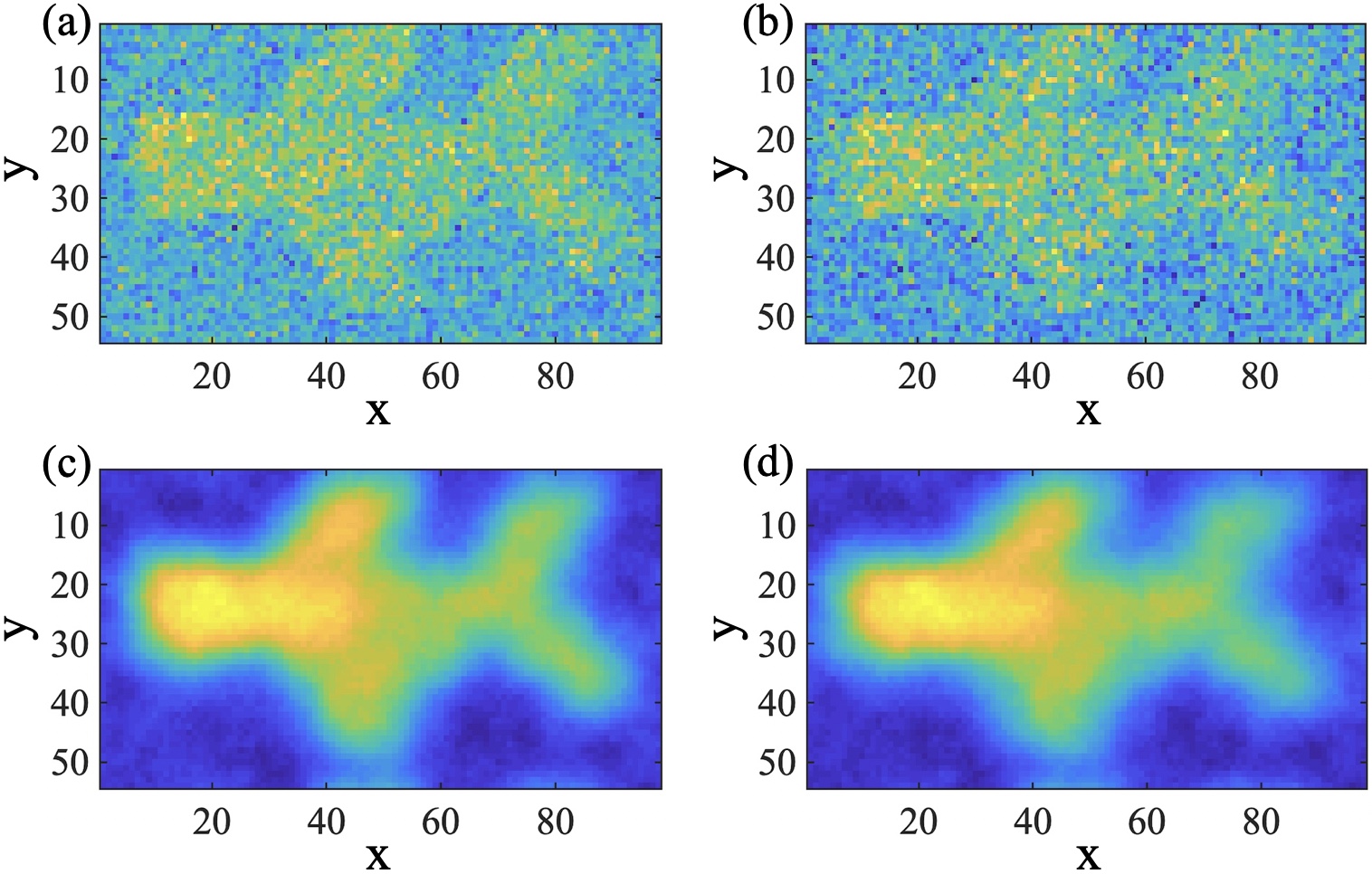}
\caption{Results by sampling 10000 patterns. (a) White noise CGI result with static target. (b) White noise CGI result with displacement = 0.9${mm}$. (c) Pink noise CGI result with static target. (d) Pink noise CGI result with displacement = 0.9${mm}$.}
   \label{fig:main}
\end{figure}
The main results are shown in Fig.~\ref{fig:main}. The results in the first column are reconstructed images with a static target, the second column has results achieved when the target moves with displacement = 0.9${mm}$. The first and second row are done with white noise and pink noise speckle patterns, respectively. It is clear that the pink noise CGI has a much higher SNR than the white noise CGI, due to the non-zero cross-correlation from the pink noise pattern.  When the target is moving, the equivalent total sampling at each position decreases significantly. Thus, the auto-correlation of the white noise is not enough to suppress environmental noise within a limited number of patterns under the circumstance of a moving object, as shown in Fig.~\ref{fig:main}(b). On the other hand, the pink noise pattern can still give a clear image of the object, as shown in Fig.~\ref{fig:main}(d).

\begin{figure*}[!htp]
\captionsetup{justification=raggedright,singlelinecheck=false}
\centering\includegraphics[width=0.75\linewidth]{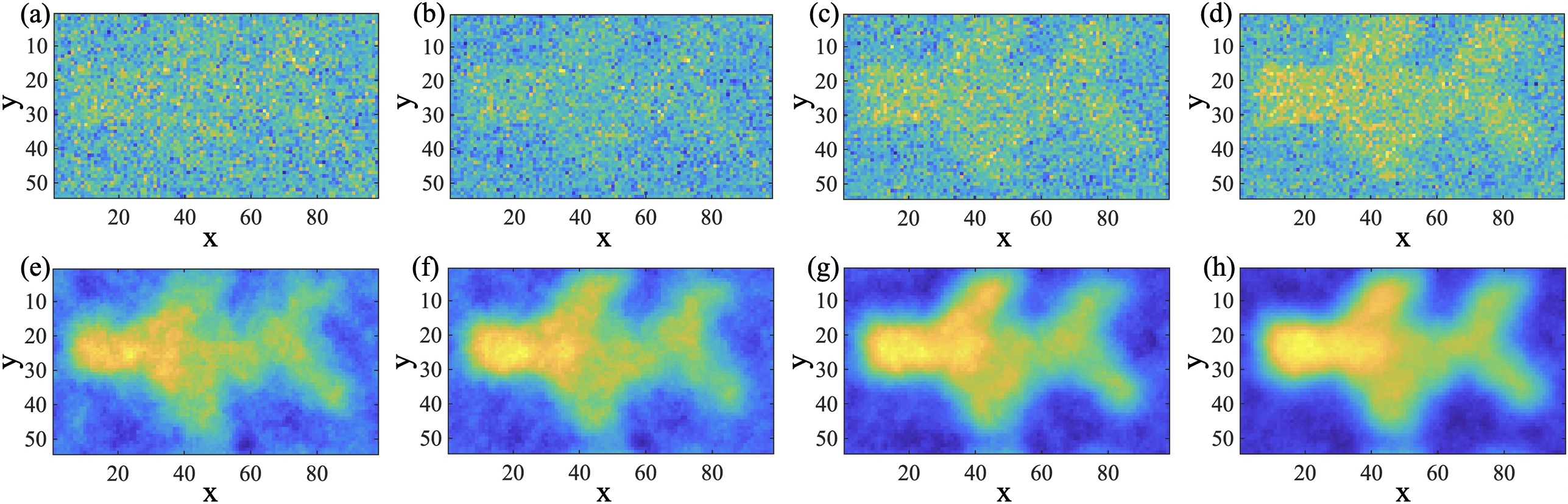}
\caption{Retrieved images with an object displacement of 0.4${mm}$. The number of sampling is 1000, 2000, 5000, and 10000 in each column, left to right. (a-d): results of white noise CGI. (e-h): results of pink noise CGI.}
   \label{fig:red&white04}
\end{figure*}
To further exam our method and explore the advantages of the pink noise speckle patterns, we also measure the target with a displacement of 0.4${mm}$ under different sample rates. The results are shown in Fig.~\ref{fig:red&white04}. The sampling number varies from 1000 to 10000. Fig.~\ref{fig:red&white04} (a-d) are white noise CGIs and Fig.~\ref{fig:red&white04} (e-h) are pink noise CGIs. From Fig.~\ref{fig:red&white04} (e) and (f), we can see that the shape of the object could be determined by pink noise CGI even the sampling number is 1000, and becomes much clearer when N gets bigger but still under Nyquist limit (5292 in our experiment). Nevertheless, the retrieved images from white noise CGI shown in Fig.~\ref{fig:red&white04} (a) and (b) are totally blurred, not clear even when ${N =5000}$ and $10000$ in Fig.~\ref{fig:red&white04} (c) and (d). The results also have poor visibility as compared with the pink noise cases. 

\begin{figure}[!htp]
\captionsetup{justification=raggedright,singlelinecheck=false}
\centering\includegraphics[width=0.55\linewidth]{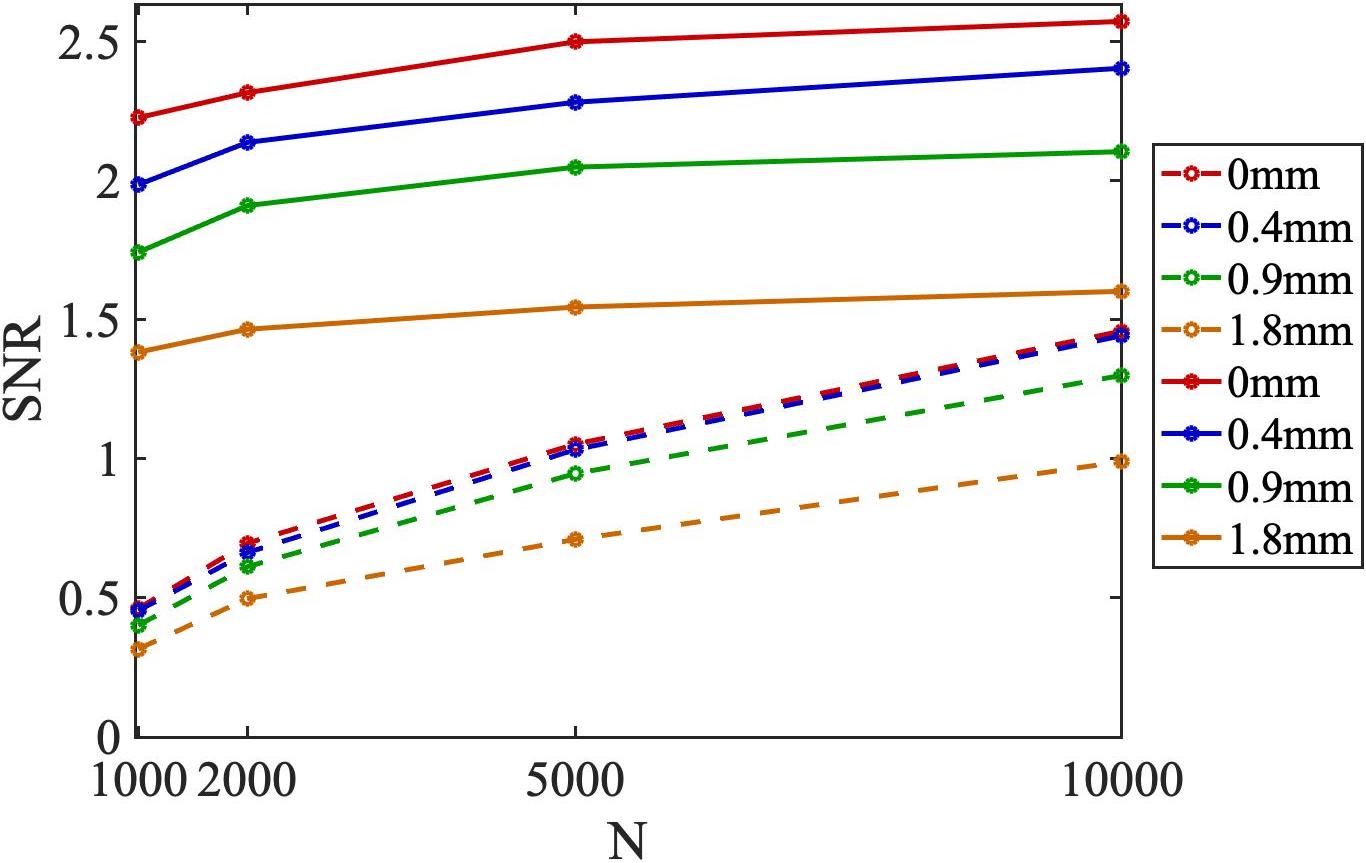}
\caption{
Signal-to-noise ratio with respect to sampling number and displacement, where pink noise CGIs are solid lines and white noise CGIs are dashed lines.}
   \label{fig:SNR_all}
\end{figure}
 Lastly, in Fig.~\ref{fig:SNR_all}, we plot the SNR of the retrieved CGIs with respect to different sampling numbers (1000, 2000, 5000, and 10000) and different displacements (0${mm}$, 0.4${mm}$, 0.9${mm}$, and 1.8${mm}$), for both white and pink noise images. The solid lines are for pink noise measurements, and dashed lines are for white noise measurements.  We see that pink noise always has a much higher SNR than white noise with measurements made under the same displacement, 4 times higher when ${N = 1000}$, for instance. For both cases, the SNR decreases when the displacement increases, as expected.

\section{Conclusion}
In this work, we have developed a practice based on the pink noise pattern in the CGI system to reconstruct images of a moving object, indicated by the simultaneously improved SNR and fewer sampling numbers. Experimentally, the SNR of pink noise CGI is usually 4 to 5 times higher than the SNR of white noise CGI, especially when the sample rate is low. Our scheme takes advantage of the positive cross-correlation, which is unique to the pink noise pattern, giving the robustness of overcoming the environmental noise, decreasing the sampling number, and further ameliorating the quality of imaging. Namely, reducing the number of patterns on DMD, directly shortening the time spent on the image reconstruction process.

We note here that, due to the cross-correlation, a low level of edge’s sharpness is the main disadvantage of pink noise CGI. Further amelioration can be done with orthonormalization method~~\cite{sun2017russian, luo2018orthonormalization} and deep learning technique~\cite{lyu2017deep,He2018Ghost}, which may also further decrease the sampling rate of the measurements. 

In conclusion, this work provides a good solution to obtain images of a moving target with unknown random speed. The strategy has broad application
prospects in target tracking, object recognition and remote sensing.

\section*{Funding}
Air Force Office of Scientific Research (Award No. FA9550-20-1-0366 DEF), Office of Naval Research (Award No. N00014-20-1-2184), Robert A. Welch Foundation (Grant No. A-1261), National Science Foundation (Grant No. PHY-2013771).
\section*{Acknowledgments}
X. N. thanks A. Svidzinsky for his kind help during his visit to IQSE, Texas A\&M University.

\section*{Declaration of competing interest} The authors declare that they have no known competing financial interests or personal relationships that could have appeared to influence the work reported in this paper.

\end{document}